\documentclass[%
reprint, twocolumn, superscriptaddress,
showpacs,preprintnumbers,
nofootinbib,
 amsmath,amssymb,
 aps,
pra,
]{revtex4}
\usepackage{mathrsfs}
\usepackage{graphicx}
\usepackage{epstopdf}
\usepackage{dcolumn}
\usepackage{bm}
\usepackage{color} 
\usepackage{CJK}
\def\la{{\langle}}
\def\ra{{\rangle}}

\newcommand{\beq}{\begin{equation}}
\newcommand{\eeq}{\end{equation}}
\newcommand{\beqa}{\begin{eqnarray}}
\newcommand{\eeqa}{\end{eqnarray}}

\newcommand{\cE}{{\cal{E}}}

\begin{document}
\title{Fast shuttling of a particle under weak spring-constant noise of the moving trap}
\author{Xiao-Jing Lu}
\affiliation{ Department of Electric and Mechatronics Engineering, Xuchang University,
Xuchang 461000, China}
\affiliation{Departamento de Qu\'{\i}mica F\'{\i}sica, UPV/EHU, Apdo.
644, 48080 Bilbao, Spain}

\author{A. Ruschhaupt}
\affiliation{Department of Physics, University College Cork, Ireland}

\author{J. G. Muga}
\affiliation{Departamento de Qu\'{\i}mica F\'{\i}sica, UPV/EHU, Apdo.
644, 48080 Bilbao, Spain}

\begin{abstract}
We investigate the excitation of a quantum particle  shuttled in a harmonic trap
with weak spring-constant colored noise.
The Ornstein-Uhlenbeck model for the noise correlation function describes
a wide range of possible noises, in particular for short
correlation times the white-noise limit examined in Lu et al, Phys. Rev. A  {89}, 063414 (2014)
and, by averaging over correlation times, ``$1/f$ flicker noise''.
We find expressions for the excitation energy in terms of static (independent of trap motion) and dynamical sensitivities,
with opposite behavior with respect to shuttling time,
and demonstrate  that  the excitation can be reduced  by  proper process timing and design of  the trap trajectory.
\end{abstract}
\maketitle
%
%
%
%
%
\section{Introduction}
Shuttling particles among specific sites without final motional excitation is
a basic operation for different fundamental studies and quantum technologies. The particles may be
single electrons \cite{Taylor2005,Golovach2010,Cadez2013,Huang2013}, single neutral atoms \cite{Steffen2012}, condensates \cite{Couvert2008},
thermal ensembles \cite{Dupont2016},
individual ions and ion chains \cite{Kielpinski2002,Rowe,Wineland2006,Mainz1,transna2,Blakestad2011,Home,Roos,Monroe,Bowler,Schmidt,An}, or large ion clouds \cite{Kamsap2015}.
Fast shuttling (compared to adiabatic transport) is in principle desirable to avoid decoherence and, in quantum information
applications with qubit transport, to speed up computation times.
Shortcuts to adiabaticity (STA) \cite{STA-review} are the set of
techniques  to design trap trajectories that achieve fast shuttling without final excitation,
in theory \cite{transport noise,Palmero2013,Torrontegui2012,Torrontegui2011,Chen2011,Mainz2,Cadez2013} and experiment \cite{Bowler,Schmidt,An,Kaufmann2018}.

Often very high fidelities with respect to target states
are needed, as in phase gates \cite{Palmero2017},
interferometers based on driving atoms along designed trajectories
\cite{Steffen2012,Dupont2016}, or
in schemes for implementing scalable architectures for quantum information processing \cite{Kielpinski2002,Rowe,Wineland2006,Home,Roos,Monroe,Taylor2005}.  Proof-of-principle experiments
in simplified settings may yield satisfactory results and demonstrate remarkable
control capabilities and bearable noise levels, see e.g.  \cite{Blakestad2011,Bowler,Schmidt,Kaufmann2018} for trapped ions,
but applications beyond prototype
level will be more demanding in terms of shuttling times, required fidelities, and travel distances.
Pushing the conditions beyond the ones for current experiments  may lead to nontrivial effects, for example the fidelities were shown to be non-monotonous with respect to the shuttling time for a white-noise in the spring constant of the driving trap, with a maximum at some shuttling  time, and decreasing for shorter  times \cite{transport noise}.
Moreover,  considering a very broad span of physical platforms and experimental settings, for which the sources and characteristics of the noise are varied and, typically, only partially known \cite{Brownnutt2015},  a fundamental understanding of the {\it effect} of noise  in the shuttling process is worthwhile and timely, considering different noise types and regimes for the values and ratios of
characteristic times involved, such as correlation times, oscillation period, and shuttling time.

Our aim here is to set a general framework for such a fundamental understanding of the effects of noise
in the spring constant of the driving trap for different noise types and regimes.
This knowledge will help to set optimal strategies to choose shuttling times and transport protocols to mitigate  the effects of noise.
When the noise characteristics are unknown the theory may serve to inversely deduce the specific regime and noise type.
Some disclaimers are in order: (i) We shall deal with single-particle shuttling transport, although some multi-particle systems may
be treated similarly,  e.g. condensates \cite{Torrontegui2012} or the center of mass of equal particles in harmonic traps \cite{Palmero2013};
(ii) We will not discuss here the sources of noise, which is of course an important topic but quite  separate from our focus;
(iii) For concreteness, we use a trapped ion in the numerical examples but the regimes explored are not limited to the ones
that might be of relevance to specific trapped-ion settings, and may as well be applicable to the shuttling of
electrons or neutral atoms.

In \cite{transport noise} the effect of several noises was studied by a Master equation valid in the limit of short
noise correlation times \cite{yuting1} compared to other characteristic times.
That equation is sufficient to analyze the effect of white-noise and perturbative corrections near that limit but cannot be used for strongly colored noise and longer correlation times.
In this paper we apply a different approach to investigate colored noise in the limit of weak noise, but for arbitrary correlation times. We consider
for that end the Ornstein-Uhlenbeck (OU) noise model. This is a useful, flexible model as we can take first the white noise (short correlation time) limit and check the consistency of the results with \cite{transport noise}; then we shall study larger correlation times. Moreover, by integrating over correlation times, the ``$1/f$ flicker noise", expected for many different noise sources  \cite{Brownnutt2015}
(where $1/f$ stands for the inverse dependence on frequency of the noise power spectrum) is
modeled without the strong restrictions in the integral limits that were required in \cite{transport noise} to stay within the validity regime of the short-correlation-time Master equation. The lower and upper integral limits for the correlation times are inversely proportional, respectively, to high-frequency and low-frequency cutoffs for the $1/f$ behavior of the noise spectrum. The dependence of the results on these limits may now be examined.

In Sec. \ref{II} we review the design of faster-than-adiabatic trap trajectories without final excitation (STA) using invariant-based inverse engineering.
In Section \ref{model}, we find  expressions for the final excitation energy with spring-constant noise using perturbation theory.
In Sec. \ref{OU}, we  discuss the Ornstein-Uhlenbeck noise model, and the (white noise) short correlation-time limit.
  Section \ref{fn} addresses  flicker noise by combining OU processes. The dependences, role, and relative importance of
``static'' and ``dynamical'' sensitivities to noise are examined. Strategies to minimize the effect of noise are proposed.
The paper ends with a summary and a technical appendix.
\section{Invariant-based inverse engineering method\label{II}}
In this section, we provide a brief review of invariant-based inverse engineering
to shuttle an ion by a moving harmonic trap with center $q_0(t)$ and time dependent
angular frequency $\omega(t)$. The Hamiltonian is
\beq
\hat{\cal H}_0 (t)=\frac{\hat{p}^2}{2 m} + \frac{1}{2}m \omega(t)^2
[\hat{q}-q_0(t)]^2.
\eeq
Subtracting the purely time-dependent term
that does not change the dynamics (it only gives a global phase to the wavefunction),
we may also use
$\hat{H}_0=\hat{\cal H}_0 (t)-m\omega(t)^2q_0(t)^2/2$,
\beq\label{vqt}
\hat{H}_0(t)=\frac{\hat{p}^2}{2 m}-F(t)\hat{q}+\frac{m}{2}\omega^2(t)\hat{q}^2,
\eeq
%
%
%
where $\hat{p}$ and $\hat{q}$ are  momentum and position
operators and $F(t)=m\omega^2(t)q_0(t)$ is a homogeneous force.

This Hamiltonian has a quadratic-in-momentum
Lewis-Riesenfeld invariant \cite{Torrontegui2011,LR,LL,DL}
\beqa
\label{inva}
\hat{I}(t)\!&=&\!\frac{1}{2m}[\rho(t)(\hat{p}-m\dot{q}_c(t))-m\dot{\rho}(t)(\hat{q}-\dot{q}_c(t))]^2
\nonumber\\
&+&\frac{1}{2}m \omega_0^2 \bigg(\frac{\hat{q}-q_c(t)}{\rho(t)}\bigg)^2,
\eeqa
which verifies
\beq
\frac{d \hat{I}(t)}{d t} \equiv \frac{\partial \hat{I}(t)}{ \partial t} +\frac{1}{i \hbar} [\hat{I}(t), \hat{H}_0(t)] =0.
\label{inva}
\eeq
provided
$\rho(t)$ and $F(t)$ satisfy the ``Ermakov'' and ``Newton'' auxiliary equations
\beqa
\ddot{\rho}(t)+\omega^2(t)\rho&=&\frac{\omega_0^2}{\rho^3(t)},
\nonumber\\
\ddot{q}_c(t)+\omega^2(t)q_c(t)&=&F(t)/m,\label{newton eq}
\label{erma}
\eeqa
with $\omega_0$ being constant.
These two equations may be found from Eq.  (\ref{inva}) assuming a quadratic-in-momentum ansatz for $\hat{I}(t)$.
$\rho(t)$ is a scaling factor that determines the width of the states
and $q_c(t)$ is a classical trajectory in the forced oscillator; it is also the center of elementary solutions
of the Schr\"odinger equation (transport modes) described below.
The eigenvalues $\lambda_n$ of $\hat{I}(t)$ are  constant,
$\hat{I}(t)\psi_n(t)=\lambda_n \psi_n(t)$, whereas the eigenstates of the invariant $\psi_n(t)$,
are time-dependent,
\beq\label{psinqt}
\psi_n(q,t)=\frac{1}{\sqrt{\rho}}e^{\frac{im}{\hbar}[\frac{\dot\rho q^2}{2\rho}
+\frac{(\dot{q}_c\rho-\dot{\rho}q_c)q}{\rho}]}\phi_n\bigg(\frac{q-q_c}{\rho}\bigg),
\eeq
where $\phi_n(q)$ are the eigenstates of a static harmonic oscillator with angular frequency $\omega_0$.
The solutions of  $i \hbar
\partial_t\Psi (q,t) = \hat{H}_0(t) \Psi (q,t)$ can be written in terms of
``transport modes'' $\Psi_n(q,t)\equiv e^{i\theta_n(t)} \psi_n(q,t)$ as
$
\Psi(q,t) = \sum_n c(n)e^{i\theta_n(t)} \psi_n(q,t),
$
where $c(n)$ are time-independent coefficients, $n=0,1,...$.
The Lewis-Riesenfeld phases $\theta_n(t)$ are  found so that  each transport mode is an exact
solution of the time-dependent Schr\"odinger equation,
\beq
\label{LRphase}
\theta_n(t) = \frac{1}{\hbar} \int_0^t \Big\langle \psi_n (t') \Big|
i \hbar \frac{\partial }{ \partial t'} - \hat{H}_0(t') \Big| \psi_n (t')  \Big\rangle d t'.
\eeq
Transport modes are orthogonal to each other at any time. They are all centered at
$q_c(t)$ and have widths proportional to $\rho(t)$.

For transport in a rigid harmonic trap two simplifications are
\beq
\omega^2(t)=\omega_0^2,~~~\rho(t)=1.
\eeq
Suppose that the harmonic trap must go from $q_0(0) =0$ to
$q_0 (T) =d$ in a shuttling time $T$. The trap trajectory $q_0 (t)$ can
be inverse engineered by designing first $q_c (t)$. To make $\hat{I}(t)$ and
$\hat{H}_0(t)$ commute at $t=0$ and $t=T$, so that they share eigenvectors
at those times
and the initial eigenvectors are dynamically mapped to final eigenvectors
by the designed transport protocol we set \cite{Torrontegui2011}
\beqa
\label{conq}
q_0(0)=q_c(0)=0,~~ \dot{q}_c(0)=0,
\nonumber
\\
 q_0(T)=q_c(T)=d,~~\dot{q}_c(T)=0.
\eeqa
In other words, with these conditions the transport modes are initially eigenstates of the initial trap
and at final time corresponding eigenstates of the final trap.
We assume continuous position functions $q_0(t)$ so the additional conditions
\beq \label{conqdd}
\ddot{q}_c(0)=0,~~\ddot{q}_c(T)=0
\eeq
are also satisfied.

%
%
%
%
%
\section{Energy sensitivities to spring-constant noise }\label{model}
We describe classical spring constant noise as
$\omega^2(t)=\omega_0^2[1+\lambda \xi(t)]$.
$\omega_0$ is the average (constant) trap frequency and $\xi(t)$ is a classical noise
that satisfies
\beq\label{con-noise}
\cE[ \xi(t)]=0,~~~\cE[ \xi(t)\xi(s)]=\alpha(t-s),
\eeq
where $\alpha(t-s)$ is the correlation function  and $\cE[\cdot\cdot\cdot]$ the statistical expectation;
$\lambda$ is the strength of the noise.
Now, as in the general case,
the  functions $\rho(t)$, which is no longer constant because of the time dependence of $\omega(t)$,
and $q_c(t)$ satisfy Eqs. (\ref{erma}).
We  assume that there is no noise at initial time, so the following
initial conditions are satisfied:
\beqa\label{in condition}
\rho(0)&=&1, ~\dot{\rho}(0)=\ddot{\rho}(0)=0,
\nonumber\\
q_c(0)&=&0,~\dot{q}_c(0)=\ddot{q}_c(0)=0.
\eeqa
A series expansion of the auxiliary functions in the noise strength $\lambda$ takes the form
\beqa\label{expansion}
\rho(t)&=&\rho^{(0)}(t)+\lambda \rho^{(1)}(t)+\cdot\cdot\cdot,
\nonumber\\
q_c(t)&=&q_c^{(0)}(t)+\lambda q_c^{(1)}(t)+\cdot\cdot\cdot.
\eeqa
We assume also that the noiseless protocol works perfectly, i.e.
the zeroth order of the auxiliary functions (noiseless limit) obeys
\beqa\label{0 order}
\rho^{(0)}(t)&=&1,\nonumber\\
\ddot{q}_c^{(0)}(t)+\omega_0^2q_c^{(0)}(t)&=&\omega_0^2q_0(t),
\eeqa
where $q_c^{(0)}(t)$ satisfies the same conditions at initial and final time as  $q_c(t)$ in Eqs. (\ref{conq}) and (\ref{conqdd}).

Combining Eqs. (\ref{erma}), (\ref{expansion}), and (\ref{0 order}), and keeping only the first order of $\lambda$, we get
\beqa\label{q1}
\ddot{\rho}^{(1)}(t)+4\omega_0^2\rho^{(1)}(t)&=&-\omega_0^2\xi(t),
\nonumber\\
\ddot{q}_c^{(1)}(t)+\omega_0^2q_c^{(1)}(t)&=&\ddot{q}_c^{(0)}(t)\xi(t),
\eeqa
with initial conditions $\rho^{(1)}(0)=\dot{\rho}^{(1)}(0)=\ddot{\rho}^{(1)}(0) $ and $q_c^{(1)}(0)=\dot{q}_c^{(1)}(0)=\ddot{q}_c^{(1)}(0) $.
The solutions of Eq. (\ref{q1}) are
\beqa\label{1 order}
\rho^{(1)}(t)&=&-\frac{\omega_0}{2}\int_0^tds\xi(s)\sin[2\omega_0(t-s)],
\nonumber\\
q_c^{(1)}(t)&=&\frac{1}{\omega_0}\int_0^tds \xi(s)\sin[\omega_0(t-s)]\ddot{q}_c^{(0)}(s).
\eeqa
We also assume that there is no noise at the final time, so the Hamiltonian at final time
is  $\hat{\cal{H}}(T)=\hat{p}^2/2m+ m\omega_0^2(\hat{q}-d)^2/2$.
The final-time energy expectation with constant frequency $\omega_0$ corresponding
to an initial state in the $n_{th}$ mode for a  realization of the noise $\xi(t)$ is exactly
calculated 
as
\beqa\label{energy}
E_{n,\xi}&=&\la \hat{{\cal H}}(T)\ra=\la \Psi_n(T)|\hat{\cal H}(T)|\Psi_n(T)\ra
\nonumber\\
&=&\frac{m}{2}\omega_0^2[q_c(T)-d]^2+\frac{\hbar\omega_0}{4}(2n+1)\frac{1+\rho^4(T)}{\rho^2(T)}
\nonumber\\
&+&\frac{m}{2}\dot{q}_c^2(T)+\frac{\hbar}{4\omega_0}(2n+1)\dot{\rho}^2(T),
\eeqa
where $\Psi_n(T)=e^{i\theta_n(T)}\psi_n(T)$, see Eq. (\ref{psinqt}).

For small noise strength $\lambda$, $E_{n,\xi}$ can be expressed as a series expansion in $\lambda$ as
\beq
E_{n,\xi}\approx E_{n,\xi}^{(0)}+\lambda E_{n,\xi}^{(1)}+\lambda^2 E_{n,\xi}^{(2)}+\cdot\cdot\cdot,
\eeq
where $E_{n,\xi}^{(1)}=\frac{\partial E_{n,\xi}}{\partial\lambda}$, $E_{n,\xi}^{(2)}=\frac{1}{2}\frac{\partial^2 E_{n,\xi}}{\partial\lambda^2}$.
From Eq. (\ref{energy}) and using the series expansion for $\rho(t)$ and $q_c(t)$ in Eq. (\ref{expansion}), we get
$E_{n,\xi}^{(0)}=\hbar\omega_0(n+\frac{1}{2})$, $E_{n,\xi}^{(1)}=0$, and
\beqa
E_{n,\xi}^{(2)}&=&m\omega_0^2q_c^{(1)}(T)^2+2\hbar\omega_0(2n+1)\rho^{(1)}(T)^2
\nonumber\\
&+&m\dot{q}_c^{(1)}(T)^2+\frac{\hbar\dot{\rho}^{(1)}(T)^2}{2\omega_0}(2n+1).
\eeqa

Now we average over different realizations of the noise and get
\beq\label{En}
E_n=\cE[ E_{n,\xi}]= E_n^{(0)}+\lambda^2 \frac{1}{2}\cE\bigg[\frac{\partial^2 E_{n,\xi}}{\partial\lambda^2}\bigg].
\eeq
where $E_n^{(0)}=E_{n,\xi}^{(0)}$ is the final energy without noise for the state $n$.
Let us define the noise sensitivity of a given transport protocol by
\beqa
G(T;n) = \frac{1}{2}\cE\bigg[\frac{\partial^2 E_{n,\xi}}{\partial\lambda^2}\bigg]=\cE[E_{n,\xi}^{(2)}].
\eeqa
Using the solution of $\rho^{(1)}(t)$ and $q_c^{(1)}(t)$ in Eqs. (\ref{1 order}) and the conditions given in Eq. (\ref{con-noise}), we find finally
\beqa
G(T;n)&=&G_1(T;n)+G_2(T;n),\nonumber\\
G_1(T)&=&\hbar\omega_0^3\bigg(n+\frac{1}{2}\bigg)\int_0^Tds\ \alpha(s)(T-s)\cos(2\omega_0s),
\nonumber\\
G_2(T)&=&\!\!m\!\int_0^T\!\!\!ds\ \alpha(s)f(s),
\label{exact-g1g2}
\eeqa
where
\begin{equation}
f(s)=\cos(\omega_0s)\!\int_0^{T-s}\!\!du\ \ddot{q}_c^{(0)}(u)\ddot{q}_c^{(0)}(u+s).
\label{fs}
\end{equation}
For a given $T$, $G_1$ is independent of the trap trajectory and therefore independent of the specific
shuttling protocol. It is also independent of the mass but it depends on $n$.
\begin{figure}[h]
\begin{center}
\scalebox{0.65}[0.65]{\includegraphics{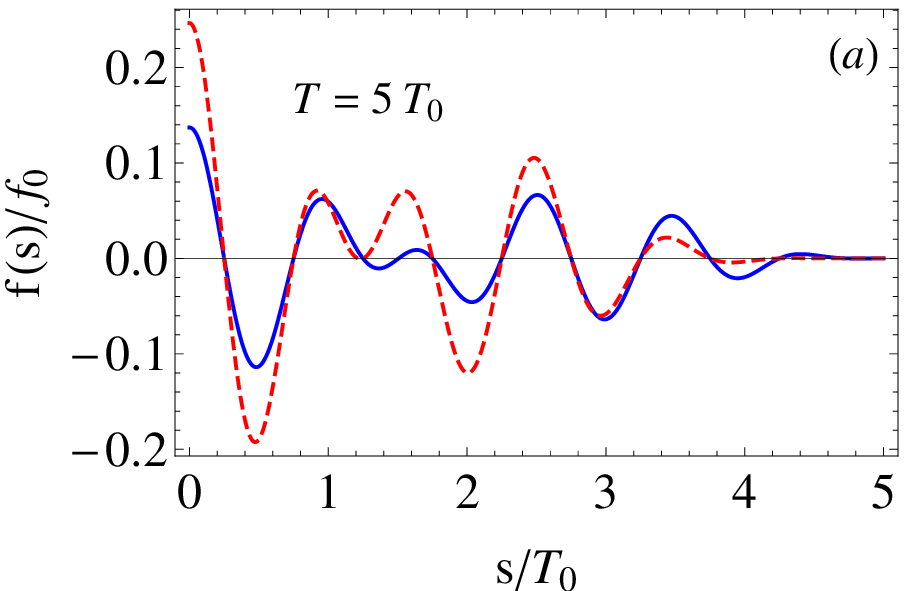}}
\scalebox{0.65}[0.65]{\includegraphics{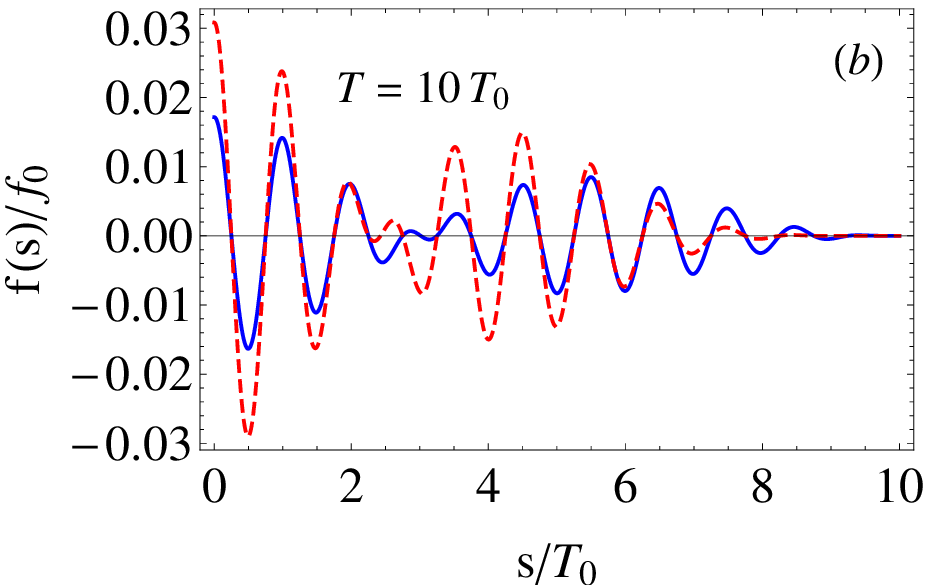}}
\caption{(Color online) The behavior of $f(s)$ for the polynomial ansatz (solid blue) and cosines protocol (dashed red).
The parameters are: $f_0=d^2/T_0^3$, $T_0=2\pi/\omega_0$, $d=280$ $\mu$m, $\omega_0=2\pi\times1.41$ MHz.}
\label{fig1-f}
\end{center}
\end{figure}
On the other hand, $G_2$ is $n$-independent, but it depends on the mass and on the chosen  $q_c^{(0)}(t)$, therefore it depends on
the trap trajectory.
We may naturally call these terms {\it static} ($G_1$) and {\it dynamical} ($G_2$) sensitivities, as the first one plays a role even for the
trap at rest, whereas the second one only appears due to the motion of the trap.
Eqs. (\ref{exact-g1g2}) and (\ref{fs}) are central results of this work that enable us to analyze the effect of weak noise
even far from the white noise limit.

To see the typical behavior of $f(s)$ in Eq. (\ref{fs}), we choose a polynomial ansatz and a cosines ansatz
for $q_c^{(0)}(t)$.
The polynomial ansatz is
\beq\label{poly}
q_c^{(0)}(t)=\sum_{n=0}^5\beta_nt^n,
\eeq
where the $\beta_n$ are determined by the conditions (\ref{conq}) and (\ref{conqdd}).
The cosines ansatz is instead given by
\beq\label{cosines}
q_c^{(0)}(t)=d\left[b_0+b_1\cos\left(\frac{\pi t}{T}\right)+b_2\cos\left(\frac{3\pi t}{T}\right)\right],
\eeq
where $b_0$, $b_1$, and $b_2$ are chosen such that the ansatz satisfies the conditions (\ref{conq}) and (\ref{conqdd}) with just three parameters.
As shown in Fig. \ref{fig1-f}, because of the factor $\cos(\omega_0s)$,
$f(s)$ oscillates and becomes zero at $s=T$, with the main peak, and most of the other peaks larger for
the cosines than for the polynomial. Also, $f(s)$ diminishes when increasing $T$ because of the
smaller accelerations involved. In particular, for the polynomial ansatz,
\beqa
f(s) &\!\!=&  \!\!\Bigg[\!\frac{720 d^2 s^7}{7T^{10}} \!\! -  \!\!\frac{360 d^2 s^5}{T^8}  \!\!+  \!\!\frac{600 d^2 s^3}{T^6} \!\! - \!\!
\frac{360 d^2 s^2}{T^5}
 \!\!+ \!\! \frac{120 d^2}{7T^3}\!\Bigg]
 \nonumber\\
&\times&\cos (\omega_0 s).
\eeqa

\subsection{Stationary traps}
To check the consistency of the theory, let us examine first the limit of stationary traps.
In several ion-trap experiments \cite{Wineland2000,SK}, the frequency dependence of the electric field noise spectral density has been
investigated by measuring heating rates for varying trap frequencies.
Specifically we focus on the spring constant noise \cite{Savard1,Savard2,Milburn}.
If the trap center does not move,
the dynamical sensitivity $G_2(T)$
is zero, as $q_c^{(0)}$ is zero at all times.  The excitation $\Delta E_n(T)=E_n(T)-E_n^{(0)}$ 
takes then the simple form ($E_n^{(0)}$ is the eigenstate energy for the state $n$)
\beqa
\Delta E_n(T)\!\!&=&\!\!\lambda^2G_1(T)\nonumber\\
\!\!&=&\!\!
E_n^{(0)} \lambda^2\omega_0^2\int_0^T \!\!dt\!\! \int_0^t \alpha(t') \cos(2\omega_0 t') dt'.
\eeqa
Thus the heating rate, for $T\gg \tau$, which enables us  to extend the upper integration limit $t$ to infinity, is
\beqa
\label{hr}
d\Delta E_n(T)/dT&=&\lambda^2dG_1(T)/dT\nonumber\\
&=&\pi\lambda^2\omega_0^2S(2\omega_0) E_n^{(0)},
\eeqa
where the spectral density of the noise (half the ``one sided-power spectrum'' in \cite{Savard2}) is
\beq
S(\Omega)=\frac{1}{\pi}\int_0^\infty \alpha(t)\cos(\Omega t) dt.
\eeq
The heating rate depends on the spectral density at the second harmonic of the trap, as found in \cite{Savard2}.
In fast non-stationary traps, however,  the term $G_2$ takes over, as we shall see,
which implies different expressions and dependences for the excitation due to noise.
%

%
%

%

%
%
%
%
\section{Ornstein-Uhlenbeck noise}\label{OU}

Now we consider Ornstein-Uhlenbeck (OU) noise.
With a finite correlation time $\tau$, the correlation function of OU noise is
\beq
\alpha(t)=\frac{1}{2\tau}e^{-t/\tau},
\eeq
with the spectrum
\beq
S(\Omega)=\frac{1}{2\pi(1+\tau^2\Omega^2)}.
\eeq
Note that in \cite{transport noise}, the noise intensity $D=\lambda^2$ (with dimensions of time) was included in the correlation function of the
Ornstein-Uhlenbeck noise.

For short correlation times $\tau/T \ll 1$. Using
the conditions $\dot q_c^{(0)} (0) = \ddot q_c^{(0)} (0) = 0$,
the expansion
$\int_0^T dt\, \frac{1}{2\tau} e^{-t/\tau} F(t) \approx \frac{1}{2} F(0) + \frac{\tau}{2} \dot F(0)
+ ...$,
and neglecting second order and higher terms, we find the sensitivities
of \cite{transport noise},
\beqa
G_1 (T;n) &=& \frac{\hbar \omega_0^3}{2} (n+1/2) \left(T - \tau + ...\right),
\\
G_2 (T;n) &=& \frac{m}{2} \int_0^T du\, \ddot q_c^{(0)} (u)^2 + ...\;.
\eeqa
As discussed in ref. \cite{transport noise}, the first term increases and the second one decreases with $T$ which implies the existence of
a minimum of sensitivity that, for common parameters
may correspond to rather large values of $T$, well within the adiabatic regime.

\begin{figure}[h]
\begin{center}
\scalebox{0.65}[0.65]{\includegraphics{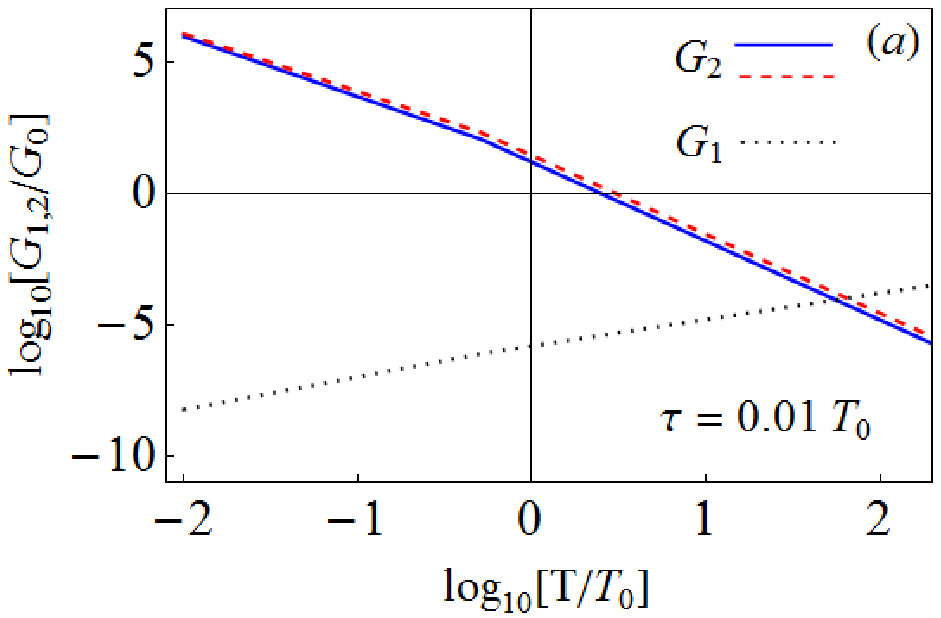}}
\scalebox{0.65}[0.65]{\includegraphics{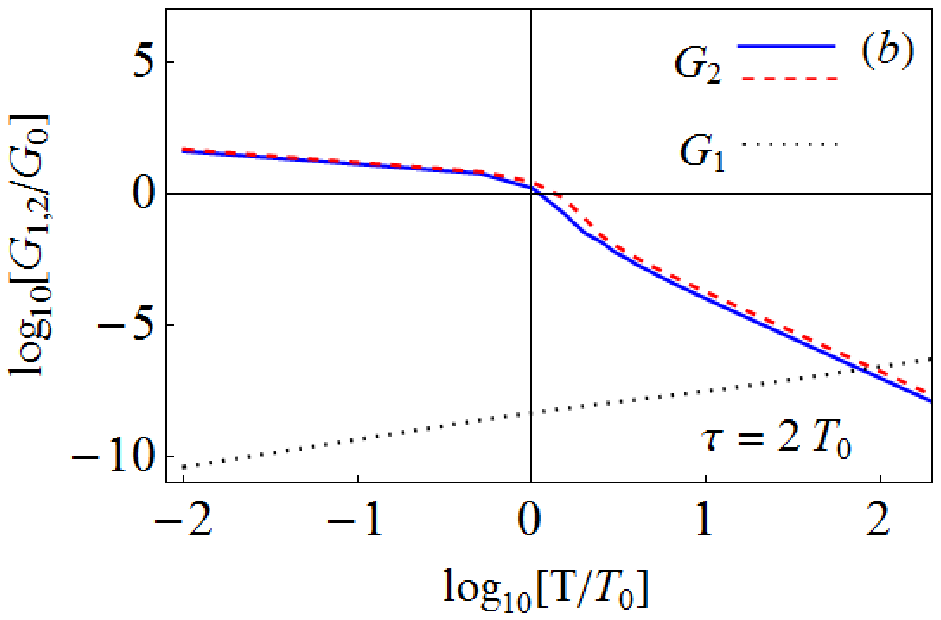}}
\caption{(Color online) $G_2$ for OU noise
versus final time $T$ using the polynomial ansatz (solid blue line) and cosines protocol (dashed red line);  $G_1$ is also shown (dotted black line). The parameters are: mass of $^{40}$Ca$^+$, initial state in $n=0$, $\omega_0=2\pi\times1.41$ MHz, $T_0=2\pi/\omega_0$, $d=280\mu$m, $G_0=10^6\hbar\omega_0^2$. 
}
\label{fig3-T}
\end{center}
\end{figure}
%

In the following we are not necessarily restricted to the limit $\tau/T\ll1$ and can also address more
general scenarios by means of Eq. (\ref{exact-g1g2}).
For OU noise, $G_1$ in Eq. (\ref{exact-g1g2}) can be calculated exactly as
\beq
G_1\!\!=\!\!\hbar\omega_0^3\bigg(n+\frac{1}{2}\bigg)\left[\frac{T}{2(1+4\tau^2\omega_0^2)}+M\right],
\label{G1exact}
\eeq
where
\beqa
M\!\!\!&=&\!\!\frac{\tau}{2(1+4\tau^2\omega_0^2)^2}\big\{(4\tau^2\omega_0^2\!\!-\!\!1)
\nonumber
\\
&-&
e^{-T/\tau}[(4\tau^2\omega_0^2\!\!-\!\!1)\!\cos(2T\omega_0)\!+\!4\tau\omega_0\sin(2T\omega_0)]\big\}.
\nonumber
\eeqa
$G_2$ is also explicit for the polynomial ansatz but much more involved, see Eq. (\ref{G2exact}) in  Appendix A.
In Fig. 2, the static $G_1$ and the dynamical $G_2$ are shown versus the final time for a $^{40}$Ca$^+$ ion, $n=0$, $d=280$ $\mu$m, and $\omega_0=2\pi\times1.41$ MHz.
We can find $G_1=G_2$ (polynomial ansatz) at  $T=55.87$ $T_0$ (the crossing point between solid line and the dotted line in Fig. \ref{fig3-T}(a)) for $\tau=0.01$ $T_0$, and $T=78.19$ $T_0$ (the crossing point between solid line and the dotted line in Fig. \ref{fig3-T}(b)) for $\tau=2$ $T_0$.
Up
to moderate values of $T$, such as $T=5 T_0$, where $T_0=2\pi/\omega_0$ is the oscillation period,
$G_1<<G_2$, so $G_1$ can typically be ignored.%

The polynomial ansatz behaves better than the cosines protocol as expected due to its  smaller $f(s)$.
In Fig. \ref{fig3-T},
the decay of $G_2(T)$
is shown for small and large $\tau/T_0$.  If $T>>\tau$ we find that $G_2\propto T^{-3}$,
consistent with a reduction of the acceleration, and thus of $f(s)$, for larger process times $T$.
This behavior is shown in the appendix, see Eqs. (\ref{G2exact}) and (\ref{g2a}).

\begin{figure}[h]
\begin{center}
\scalebox{0.65}[0.65]{\includegraphics{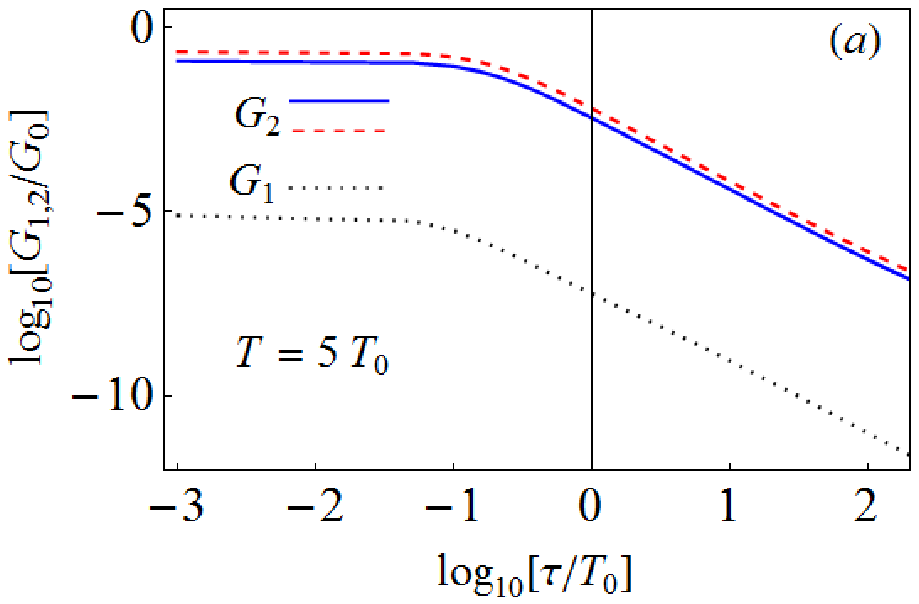}}
\scalebox{0.65}[0.65]{\includegraphics{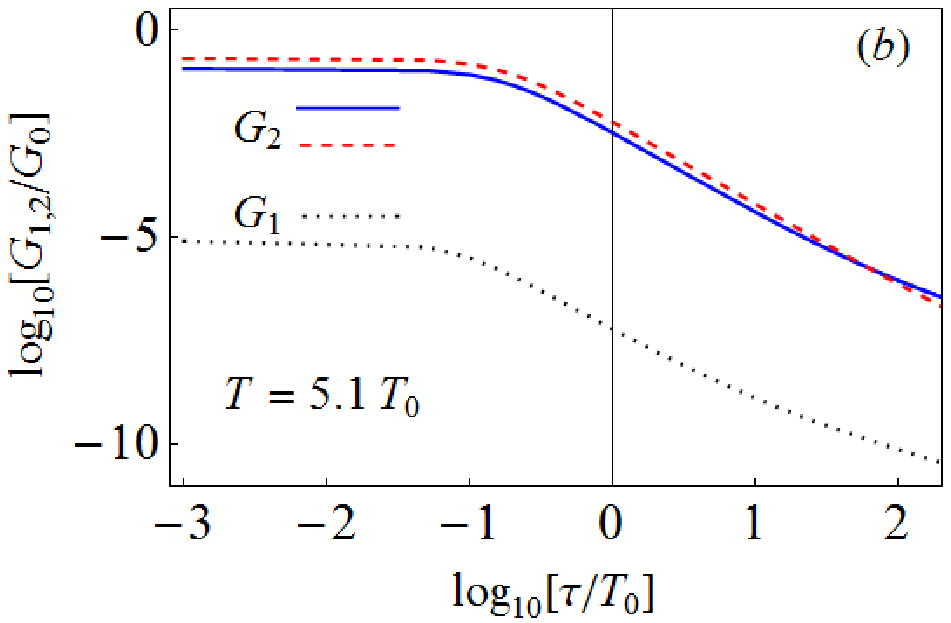}}
\caption{(Color online) $G_2$ for OU noise
versus correlation time $\tau$ using the polynomial ansatz (solid blue line) and cosines protocol (dashed red line);  $G_1$ versus $\tau$ (dotted black line). 
Other parameters are the same as in Fig. \ref{fig3-T}. }
\label{fig4-tau}
\end{center}
\end{figure}

\begin{figure}[h]
\begin{center}
\scalebox{0.65}[0.65]{\includegraphics{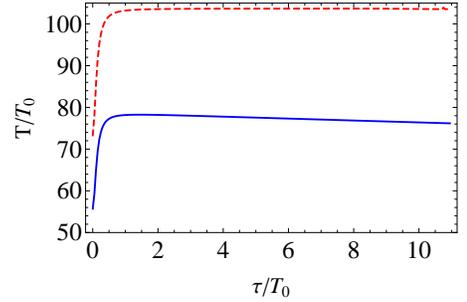}}
\caption{(Color online) The solid blue line represents the points in the $T,\tau$ plane for which $G_1=G_2$ using the polynomial ansatz. $G_2>G_1$ below the line. For a given $\tau$, the dashed red line represents the value of $T$ where $G_1+G_2$ is minimum. Mass of $^{40}$Ca$^+$, $n=0$, $\omega_0=2\pi\times1.41$ MHz, $T_0=2\pi/\omega_0$, $d=280$ $\mu$m.}
\label{T-tau}
\end{center}
\end{figure}
For $T\gg\tau\gtrsim T_0/(4\pi)$, using $e^{-T/\tau}\approx0$ and $1+4\tau^2\omega_0^2\approx4\tau^2\omega_0^2$, the expressions in Eq. (\ref{G1exact}) and (\ref{G2exact}) can be approximated as
\beq\label{middle-g1g2}
G_1\approx\frac{\hbar\omega(T+\tau)}{8\tau^2}\bigg(n+\frac{1}{2}\bigg),~~
G_2\approx\frac{60md^2}{7T^3\omega^2\tau^2}.
\eeq
Both $G_{1,2}\propto\tau^{-2}$ when $T\gg\tau\gtrsim T_0/(4\pi)$.
For larger $\tau$, so that
$e^{-T/\tau}\approx1-T/\tau$, $G_1$ is approximated as
\beqa
G_1&\approx&\hbar\omega_0^3\bigg(n+\frac{1}{2}\bigg)\bigg[\frac{T\cos^2(\omega_0T)}{4\tau^2\omega_0^2}+\frac{\tau\sin^2(\omega_0T)}{4\tau^2\omega_0^2}\nonumber\\
&-&\frac{T_0/(2\pi)}{4\tau^2\omega_0^2}\sin(\omega_0T)\cos(\omega_0T)\bigg].
\eeqa
For   $\tau>>T$, we find that
in general
$G_1\propto \tau^{-1}$  except for some special cases: when $\omega_0T=N\pi, N=0,\pm1,\pm2...$, $\sin(\omega_0T)=0$ and $\cos(\omega_0T)=\pm1$, so $G_1\propto\tau^{-2}$. In Appendix \ref{analytical-G2}, we give the expression of $G_2$, which decays as $\tau^{-1}$ when $\tau\rightarrow \infty$.

Fig. \ref{fig4-tau} demonstrates the dependence of the sensitivities with respect to $\tau$.
In Fig. \ref{fig4-tau}(a), for time $T=5T_0$, $G_2(\tau)$ tends to a constant value $mf(0)/2$
for small $\tau$ ($\tau<<T_0$), decays as
$G_2\propto\tau^{-2}$ for $\tau\gtrsim T_0/(4\pi)$; the transition to $G_2\propto\tau^{-1}$ cannot be seen in the scale of the figure.
$G_1(\tau)$ tends to $\hbar\omega^3T/4$
for small $\tau$ ($\tau<<T_0$) and  $G_1\propto\tau^{-2}$  for $\tau\gtrsim T_0/(4\pi)$.
The transition to $\tau^{-1}$ decay is shown for $G_1$ and $G_2$ in Fig. \ref{fig4-tau}(b), for a different shuttling time $T=5.1T_0$ ($\omega_0T\neq2N\pi$),
when $\tau>T$.

The curve in the $T,\tau$ plane where $G_1=G_2$ is shown in
Fig. \ref{T-tau} (solid blue line) for $n=0$ and the polynomial ansatz.
For $\tau>T_0$, $\tau<<T$ along the curve so we may use $G_1$ and $G_2$ in Eq. (\ref{middle-g1g2}), to describe
the curve approximately as
\beq
T+\tau=\frac{a}{T^3},
\eeq
where $a=960md^2/(7\hbar\omega^3)$. This curve is important as it marks the transition between
regimes dominated by static or dynamical sensitivities.

Similarly, the value of $T$ where $G=G_1+G_2$ is minimum, for fixed $\tau$,
 dashed red line in Fig. \ref{T-tau}, goes to a constant $T=\sqrt[4]{\frac{2880md^2}{7\hbar\omega^3}}$ for $\tau>T_0$,
which again may be found from the approximate $G_1$ and $G_2$ in Eq. (\ref{middle-g1g2}).

\begin{figure}[h]
\begin{center}
\scalebox{0.6}[0.6]{\includegraphics{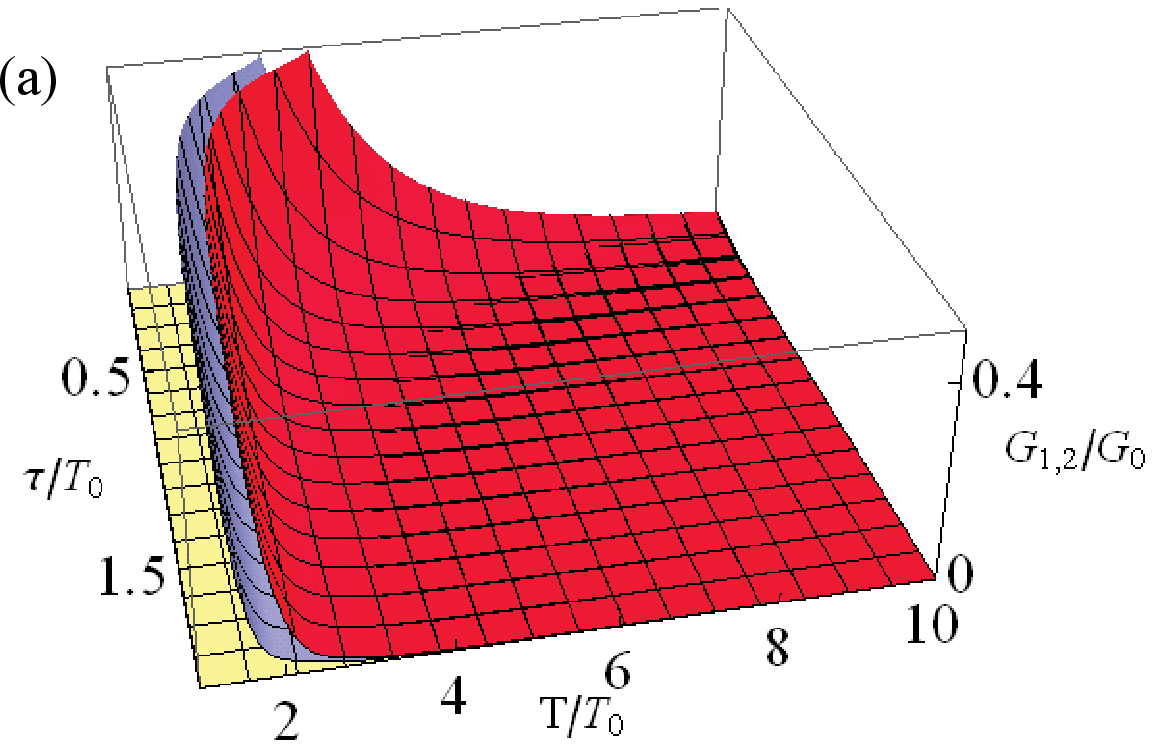}}
\scalebox{0.6}[0.5]{\includegraphics{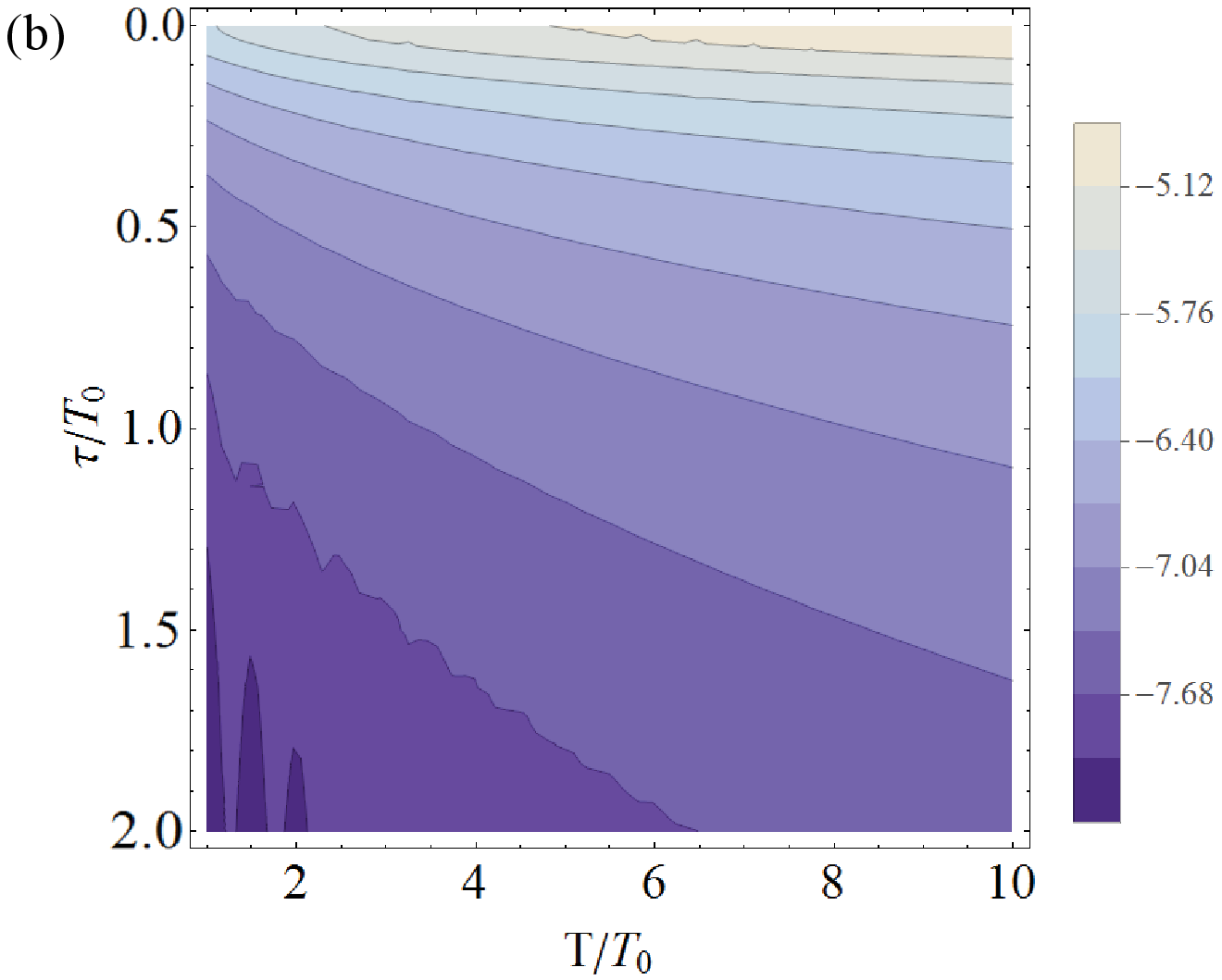}}
\scalebox{0.6}[0.5]{\includegraphics{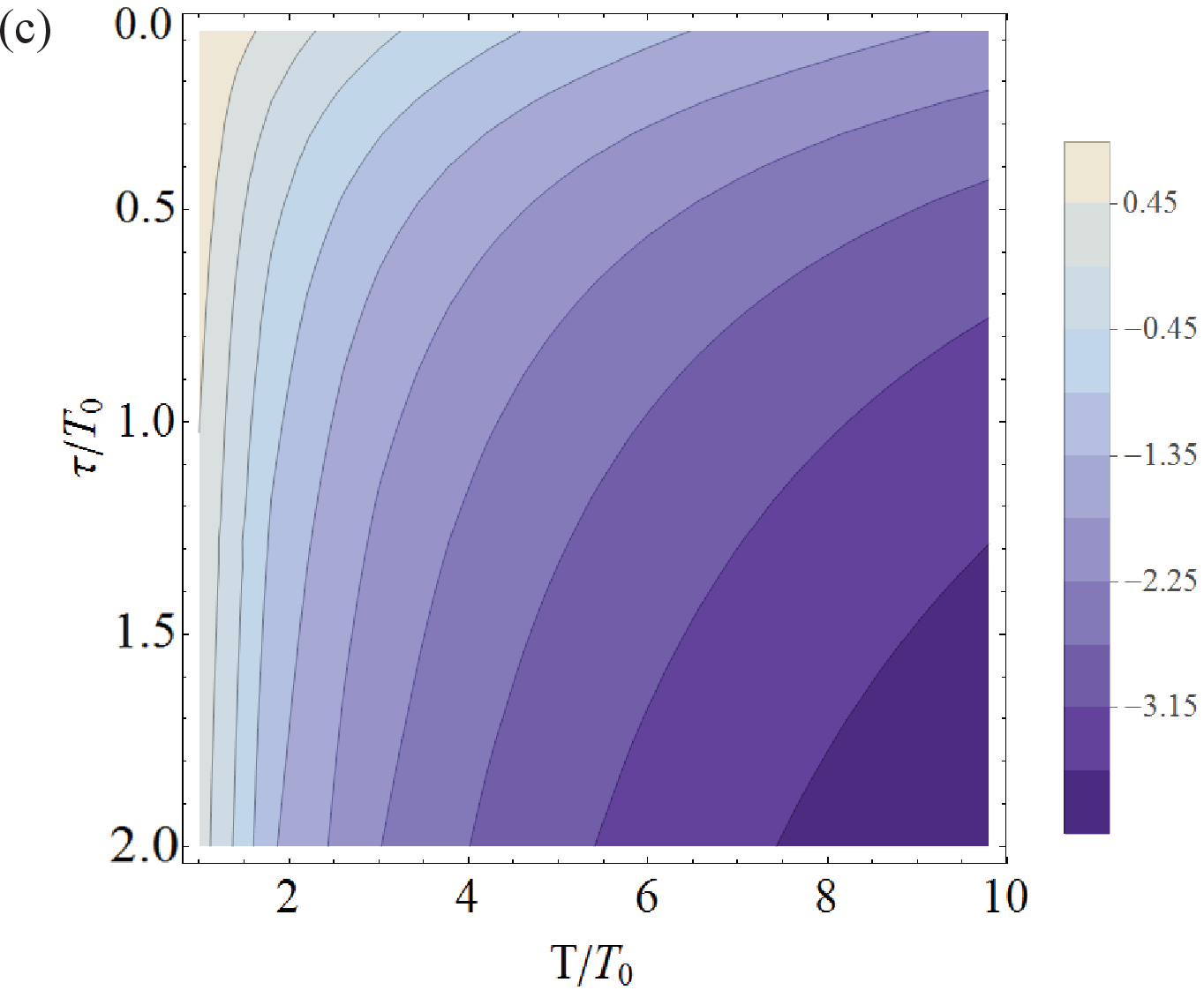}}
\caption{(Color online)(a) Sensitivities for OU noise: (a) $G_{2}$
versus correlation time $\tau$ and final time $T$ using the polynomial ansatz (middle blue surface) and cosines protocol (upper red surface);  $G_1$ (lower yellow
surface). (b) Contourplot of $log_{10}G_1$ versus $T$ and $\tau$.
(c) Contourplot of $log_{10}G_2$ versus $T$ and $\tau$.
The parameters are the same as in Fig. \ref{fig3-T}.
}
\label{fig5}
\end{center}
\end{figure}

%

Fig. \ref{fig5} combines the dependences of $G_1$ and $G_2$ on $T$ and $\tau$. The polynomial protocol outperforms the cosines protocol in the range of $T$ and $\tau$ shown.  $G_2>G_1$ in the domain of the figure so the dynamical sensitivity dominates.
The figure suggests  a possible strategy, namely to make $T$ just large enough,
a few oscillation periods $NT_0$, to be in the plateau (in the linear plot (a)) and make $G_2$ negligible. The value of $N$ depends on $\tau$ and may be
easily estimated numerically.
$G_1$ increases with $T$ for a given $\tau$, see the logarithmic plot in (b), but it is still small for the moderate values of $T$ in the plateau area of figure (a).
A comparison with Fig. \ref{T-tau} demonstrates that there is ample room for implementing fast protocols where the
effect of noise can be strongly suppressed due to small sensitivities  $G_1$ and $G_2$.

We have also tried to reduce the sensitivity by adding one more parameter in the polynomial ansatz,
\beq\label{high-poly}
q_c^{(0)}(t)=\sum_{k=0}^6n_kt^k.
\eeq
%
Here seven parameters $n_k$ are determined by six conditions (\ref{conq}) and (\ref{conqdd}), so
one of them,  $n_6$, is left free to minimize the sensitivity $G_2$.
We find numerically that for all values of $T$ and $\tau$ in Fig. \ref{fig5},
$n_6=0$ minimizes $G_2$
so that the fifth order polynomial (\ref{poly}) is in fact optimal in the chosen function space.
The use of a more general ansatz or Optimal Control Theory may be of interest but is not
pursued here.

\section{Flicker noise\label{fn}}

Flicker  ($1/f$) noise is a widespread type of colored noise.
The correlation function of flicker noise we consider here \cite{Hooge,watanabe} is
the result of averaging over OU noises with different correlation times in a range
$[\tau_1, \tau_2]$,
\beqa
\label{alfli}
\alpha_f(t)&=&\frac{1}{\tau_2 - \tau_1}\int_{\tau_1}^{\tau_2}d\tau\,\frac{1}{2\tau}e^{-t/\tau}\nonumber\\
&=&\frac{\alpha_f(0)}{\ln(\tau_2/\tau_1)}\int_{\tau_1}^{\tau_2}d\tau\,\frac{1}{\tau}e^{-t/\tau}
\nonumber\\
&=&\frac{Ei(-t/\tau_1)-Ei(-t/\tau_2)}{2 (\tau_2 - \tau_1)},
\eeqa
where $\alpha_f(0) = \frac{\ln(\tau_2/\tau_1)}{2(\tau_2-\tau_1)}$, and
$Ei(-x)=\int_{-\infty}^{-x}(e^t/t)dt$ (for $x>0$).
(In \cite{transport noise}, $\lambda^2$ was included in the correlation function.)

Determining the effective frequency cutoffs in current ion-transport experiments is  an
open question that may  depend highly on the setting \cite{lowfcut,Brownnutt2015}.

The corresponding power spectrum takes the form
\beqa
S(\Omega) &=& \frac{1}{\tau_2 - \tau_1} \int_{\tau_1}^{\tau_2} d\tau\, \frac{1}{2\pi} \frac{1}{1+\Omega^2 \tau^2}
\nonumber\\
&=&\frac{\cot^{-1}(\tau_1\Omega)-\cot^{-1}(\tau_2\Omega)}{2\pi(\tau_2-\tau_1)\Omega},
\eeqa
so that $S(\Omega)\sim 1/\Omega$ between
low-frequency and high-frequency cutoffs $\Omega_2=2\pi/\tau_2$ and $\Omega_1=2\pi/\tau_1$,
and behaves as $\sim1/\Omega^2$ beyond $\Omega_1$,  see \cite{transport noise}.

Figs. \ref{fig4-tau} and \ref{fig5} give a good hint on what to expect for the dynamical sensitivity $G_2$ with flicker noise.
By averaging over OU noises from $\tau_1$ to $\tau_2$, the flicker-noise dynamical sensitivity $G_2$ will be reduced with respect to the value for OU noise at the smaller limiting time, $G_2(\tau_1; OU)$,
when increasing $\tau_2$. In other words, the OU-noise dynamical sensitivity at a given correlation time $\tau_1$ sets a bound for the flicker-noise sensitivity when averaging for larger correlation times.
Let us now analyze some limiting cases: The  power spectrum converges for  $\tau_1 \to 0$ and $\tau_2 \to 0$ to the spectrum of white noise, i.e.
$S(\Omega) \to 1/(2 \pi)$.
In the limit $\tau_2 \to \infty$ with fixed $\tau_1$ or $\tau_1\to 0$,
the spectrum becomes zero, i.e. $S(\Omega) \to 0$ for $\Omega \neq 0$ so the effect of noise vanishes.

Using the asymptotic behaviour $Ei(-x)\simeq\gamma_E+\ln(x)-x$ for $x\rightarrow0$ and $Ei(-x)\simeq-e^{-x}/x$ for $x\rightarrow\infty$, where $\gamma_E$ is Euler's constant,
we may also analyze the case $\tau_2\gg T$ and $\tau_1\gg T$,
for which
\beqa
\alpha_f(t) \approx \frac{\ln(\tau_2/\tau_1)}{2(\tau_2-\tau_1)} - \frac{t}{2\tau_1\tau_2}.
\eeqa
We can neglect for $t\ll\tau_1$
the second term, i.e.
$\alpha_f(t) \approx \frac{\ln(\tau_2/\tau_1)}{2(\tau_2-\tau_1)} = \alpha_f(0)$
and $G_{1,2}$ are approximated as
\beqa\label{G2-f}
G_1(T)\!\!&=&\!\!\alpha_f(0)\frac{\hbar\omega_0}{2}\bigg(n+\frac{1}{2}\bigg)\sin^2(\omega_0 T),
\nonumber\\
G_2(T)\!\!&=&\alpha_f(0)m\int_0^T\!\!ds f(s).
\eeqa
\begin{figure}[h]
\begin{center}
\scalebox{0.65}[0.65]{\includegraphics{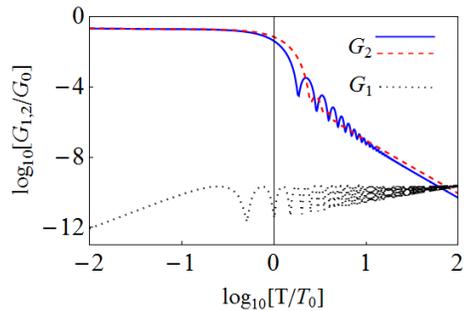}}
\caption{(Color online)  $G_2$ for flicker noise using the polynominal ansatz (solid blue line) and cosines ansatz (dashed red line), and $G_1$ (dotted black line) versus $T$. $\tau_1=80T_0$, $\tau_2=100T_0$, and other parameters are the same as in Fig. \ref{fig3-T}.}
\label{fig-fT}
\end{center}
\end{figure}
Interestingly $G_1(T)$ vanishes at every half oscillation period, $T=n\pi/\omega_0$. In Fig. \ref{fig-fT}, we plot $G_1$ (dotted line) using Eq. (\ref{exact-g1g2}), so the minima at $T=n\pi/\omega_0$ are not exactly zero.
The maxima take the value $\alpha_f(0)\hbar\omega_0/4$, which for parameters in Fig. \ref{fig-fT},
is negligible compared to $G_2(T)$  for moderate values of
$T>T_0$. This value of the maxima holds approximately for $T<\tau_1$.

For the polynomial ansatz, we get the analytical expression for $G_2$ in Eq. (\ref{G2-f}),
\beq\label{G2-ff}
G_2\!\!=\!\!\frac{7200\alpha_f(0)md^2}{\omega_0^8 T^{10}}\bigg[\!6\omega_0T\cos\!\bigg(\!\frac{u}{2}\!\bigg)+(\omega_0^2 T^2-12)\!\sin\!\bigg(\!\frac{u}{2}\!\bigg)\!\bigg]^2,
\eeq
where $u=\omega_0 T$.
When $T\ll T_0$,
$G_2\approx \alpha_f(0)md^2\omega^2/2$, as shown in Fig. \ref{fig-fT}.

For a fixed $\tau_1$, the important dependence of $G_2$ is thus in $\alpha_f(0)$ which decreases monotonously
towards zero when  increasing $\tau_2$.
In the limit $\tau_1\to\tau_2$, we have $\alpha_f(0)\to1/(2\tau_2)$, $G_2\propto\tau_2^{-1}$ , and this result converges to the large
$\tau$ limit of $G_2$ for OU noise in Eq. (\ref{G2-ou-tau}).

\section{Summary\label{discu}}
Research on noise and its effects on the control of quantum systems is an active field for fundamental and practical reasons.
Here we present a theory to understand and possibly control the effect of spring-constant noise in the shuttling of a quantum particle driven by a moving
harmonic trap.  Shuttling is an important operation for many different systems (e.g. electrons, ions, neutral atoms, or condensates) and applications
such as interferometry or quantum information processing,
so that a generic (weak noise) theory is worked out without focusing on specific systems, for which many different
noise types and noise sources, typically not fully understood and experiment-dependent, may occur \cite{Brownnutt2015}.
Such a theory is intended as a useful guiding aid for the plethora of possible scenarios.

The calculations are done for a trapped ion for illustration purposes but all possible regimes for ratios among characteristic
times are discussed.
Applying a perturbative treatment for weak noise, explicit expressions for the excitation energy are found,
which are valid for strongly colored noise and also for $1/f$ noise. Dynamical and static contributions
with opposite behavior with respect to the shuttling time are identified. Very short shuttling times are not the best option,
as the dynamical contribution increases for decreasing shuttling times.  In this sense,
the problems with noise are not necessarily solved even if current experiments
for some of the mentioned systems and given shuttling times achieve reasonable fidelities and noise mitigation.
The effect of noise may easily reappear for faster processes due to the dynamical term.
In this work the trap trajectories are designed by shortcut-to-adiabaticity (STA)
techniques so that in the noiseless limit the particle is not excited at destination with respect to the initial energy.

The Ornstein-Uhlenbeck (OU) noise plays a central role as it can describe white noise for short correlation times $\tau$
but also colored noise and
$1/f$ flicker noise by averaging over the correlation times between short and large correlation-time cutoffs.
We recover the results in the white noise limit which were found in a previous paper
\cite{transport noise} but also go beyond that limit.
The effect of OU noise on the final energy excitation is characterized in detail, analytically for specific trap trajectories. Because of the averaging implied in  flicker noise, it  may be regarded as a weakened version of the OU noise for the smaller time-cutoff.

We have also investigated the strategies and trajectories to reduce the excitation due to noise.
The main effect is achieved by a proper choice of the shuttling time,
whereas the effect of the particular trap trajectory is smaller in comparison.
STA approaches offer typically a family of possible protocols for a given process time.
Here we have only optimized the trajectory within a limited subset of smooth functions,
a full optimization with optimal control theory  with respect of the trajectory \cite{Chen2011}
remains an open question.

Finally, particle shuttling is just one among the set of motional operations currently considered
to develop quantum technologies, and specifically quantum information processing.
For trapped ions, in particular, progress has been done
to implement operations such as fast separation/recombination of multi-ion chains,
possibly with different species, expansions/compressions, or rotations.
The findings of the present work provides a basis to investigate the effects of noise in these operations.
Similarly the implementation of STA techniques advanced from one-particle shuttling \cite{Torrontegui2011}
to other operations \cite{Palmero2013,Palmero2014,Lu2015,Palmero2015a,Palmero2015b,Palmero2016,Palmero2017}
followed that step-by-step scheme.

{\it{Acknowledgments}---}.
We thank Joseba Alonso for an illuminating discussion.
This work was supported by National Natural Science Foundation of China (Grant No. 11747144), XuChang University (Grant No. 2018ZD007), the Basque Country Government (Grant No. IT472-10), Ministerio de Econom\'\i a y Competitividad (Grant No. FIS2012-36673-C03-01), and the program UFI 11/55 of the Basque Country University.\\

\appendix
\section{Analytical $G_2$ for OU noise}\label{analytical-G2}
For OU noise, the exact $G_2$ using the polynomial ansatz can be calculated
as
\beqa
G_2&=&\frac{60m d^2}{7T^{3}(1+X^2)^8}
\big\{M_1
\nonumber\\
&+&\!e^{-T/\tau}[M_2\cos(\omega_0 T)\!+\!
M_3\sin(\omega_0 T)]\big\},
\label{G2exact}
\eeqa
where $M_1$, $M_2$, $M_3$ are polynomials in $\tau/T$ and
$X=\omega_0\tau$,
\begin{widetext}
\beqa
M_1&=&(\tau/T)^7(30240\!-\!846720X^2\!\!+\!\!2116800X^4\!\!-\!\!846720X^6+30240X^8)\!\!\nonumber\\
&+&(\tau/T)^5(\!-\!2520\!+\!32760X^2\!+\!35280X^4\!-\!35280X^6\!\!-\!\!32760X^8\!\!+\!\!2520X^{10})\nonumber\\
&+&(\tau/T)^3(210-420X^2-3570X^4-5880X^6-3570X^8-420X^{10}+210X^{12})\nonumber\\
&+&(\tau/T)^2(-42-84X^2+210X^4+840X^6+1050X^8+588X^{10}+126X^{12})\nonumber\\
&+&\big(1+7X^2+21X^4+35X^6+35X^8+21X^{10}+7X^{12}+X^{14}\big),
\\
M_2&=&210\big[(\tau/T)^7(-144+4032 X^2-10080 X^4+4032 X^6-144X^8)\nonumber\\
&+&(\tau/T)^6(-144+2880 X^2-2016 X^4-4032 X^6+1008 X^8)\nonumber\\
&+&(\tau/T)^5(-60+780 X^2+840 X^4-840 X^6-780 X^8+60X^{10})\nonumber\\
&+&(\tau/T)^4(-12+84 X^2+264 X^4+168 X^6-60 X^8-60X^{10})\nonumber\\
&+&(\tau/T)^3(-1+2 X^2+17 X^4+28 X^6+17 X^8+2X^{10}-X^{12})\big],
\\
M_3&=&840X\big[(\tau/T)^7(288-2016 X^2+2016 X^4-288 X^6)\nonumber\\
&+&(\tau/T)^6(252-1008 X^2-504 X^4+720 X^6-36 X^8)\nonumber\\
&+&(\tau/T)^5(90-120 X^2-420 X^4-120 X^6+90 X^8)\nonumber\\
&+&(\tau/T)^4(15+15 X^2-42 X^4-66 X^6-21 X^8+3X^{10})\nonumber\\
&+&(\tau/T)^3(1+3 X^2+2 X^4-2X^6-3 X^8-X^{10})\big],
\eeqa
\end{widetext}
In the limit that $\tau\ll T$, $e^{-T/\tau}\rightarrow0$ and $\tau/T\ll 1$, so
\beq\label{g2a}
G_2\approx 60md^2/(7T^3),
\eeq
as shown in Fig. \ref{fig3-T}.
As $\tau\sim\infty$, $G_2$ can be approximated as

\beqa\label{G2-ou-tau}
G_2\!\!&\approx&\!\!\frac{1800md^2}{T^{10}\omega^8\tau}\bigg[144+12u^2+u^4
\nonumber\\
\!\!&-&\!\!(144\!\!-\!\!60u^2\!+\!u^4)\cos u
\!\!-\!\!(144u\!\!-\!\!12u^3)\sin \!u\bigg]\nonumber\\
&=&\frac{3600md^2}{T^{10}\omega^8\tau}\bigg[6u\cos\bigg(\frac{u}{2}\bigg)+(u^2-12)\sin\bigg(\frac{u}{2}\bigg)\bigg]^2,\nonumber\\
\eeqa
where $u=\omega_0 T$.
%


\end{document}